\documentclass{ifacconf}

\usepackage{graphicx}      
\usepackage{natbib}        

\usepackage{amsmath,amssymb,amsfonts}
\usepackage{textcomp}
\usepackage{array}
\usepackage{makecell}

\usepackage[utf8]{inputenc}
\usepackage{graphics} 
\usepackage{epsfig} 
\usepackage{times} 

\usepackage{graphicx}      
\usepackage{siunitx}
\usepackage{pgfplots,filecontents}
\usepackage{tikz,amsmath}
\usetikzlibrary{external}
\usepackage{tikzscale}
\usepackage{algpseudocode}
\usepackage{algorithm}
\usepackage{fancyhdr}
\usepackage{subcaption} 
\usepackage{eso-pic}
\usepackage{mathdots}
\usepackage{accents}

\usetikzlibrary{shapes,arrows}
\usetikzlibrary{positioning}
\usetikzlibrary{calc,patterns,
	decorations.pathmorphing,
	decorations.markings}

\usepackage{enumitem}

\pgfplotsset{compat=newest}
\pgfplotsset{plot coordinates/math parser=false}
\newlength\figureheight
\newlength\figurewidth

\newlength\defcolwidth

\usepackage{tikz}
\usepackage{tikzscale}

\definecolor{myorange}{cmyk}{0,0.35,0.85,0} 
\definecolor{mypurple}{cmyk}{0.5,1,0,0} 

\definecolor{matblue1}{rgb}{0,0.4470,0.7410}
\definecolor{matred1}{rgb}{0.85,0.325,0.098}
\definecolor{matyel1}{rgb}{0.9290, 0.6940, 0.1250}
\definecolor{matpur1}{rgb}{0.4940, 0.1840, 0.5560}
\definecolor{matgre1}{rgb}{0.4660, 0.6740, 0.1880}
\definecolor{matblue2}{rgb}{0.3010, 0.7450, 0.9330}
\definecolor{matred2}{rgb}{0.6350, 0.0780, 0.1840}
\definecolor{matgrey1}{rgb}{0.5, 0.6, 0.7}
\definecolor{matpink1}{rgb}{1, 0.07, 0.65}
\definecolor{matblue3}{rgb}{0.07, 0.62, 1}

\newcommand{\reddash}{\raisebox{2pt}{\tikz{\draw[-,matred1,dashed,line width = 0.9pt](0,0) -- (3mm,0);}}}

\newcommand{\blueline}{\raisebox{2pt}{\tikz{\draw[-,matblue1,solid,line width = 0.9pt](0,0) -- (3mm,0);}}}

\newcommand{\bluedot}{\raisebox{.7pt}{\tikz{\draw[-,matblue1,solid,line width = 1pt](0,0) circle (.7mm);}}}

\graphicspath{{./Figures/}}

\newcommand*{\tran}{^{\mkern-1.5mu\mathsf{T}}}

\graphicspath{{../Figures/}}

\theoremstyle{definition}

\begin{document}
	\AddToShipoutPictureBG*{%
		\AtPageUpperLeft{%
			\setlength\unitlength{1in}%
			\hspace*{\dimexpr0.5\paperwidth\relax}
			\makebox(0,-2)[c]{
				\parbox{\paperwidth}{ \centering
					Leontine Aarnoudse and Tom Oomen, Automated MIMO Motion Feedforward Control:\\ Efficient Learning through Data-Driven Gradients via Adjoint Experiments and Stochastic Approximation, \\
					To appear in {\em Modeling, Estimation and Control Conference 2022}, Jersey City, New Jersey, USA, 2022}}%
	}}
\begin{frontmatter}

\title{Automated MIMO Motion Feedforward Control: Efficient Learning through Data-Driven Gradients via Adjoint Experiments and Stochastic Approximation\thanksref{footnoteinfo}} 

\thanks[footnoteinfo]{This work is part of the research programme VIDI with project number
	15698, which is (partly) financed by the NWO.}

\author[TUe]{Leontine Aarnoudse}  
\author[TUe,TUd]{Tom Oomen}

\address[TUe]{Dept. of Mechanical Engineering, Control Systems Technology, Eindhoven University of Technology, Eindhoven, The Netherlands (e-mail: l.i.m.aarnoudse@tue.nl)}  
\address[TUd]{Delft Center for Systems and Control, Delft University of Technology, Delft, The Netherlands} 

\begin{abstract}                

Parameterized feedforward control is at the basis of many successful control applications with varying references. The aim of this paper is to develop an efficient data-driven approach to learn the feedforward parameters for MIMO systems. To this end, a cost criterion is minimized using a stochastic gradient descent algorithm, in which both the search direction and step size are determined through system experiments. In particular, the search direction is chosen as an unbiased estimate of the gradient which is obtained from a single experiment, regardless of the size of the MIMO system. The approach is illustrated using a simulation example, in which it is shown to be superior to a deterministic method in terms of convergence speed and thus experimental cost.   

\end{abstract}


\end{frontmatter}

\setlength\defcolwidth{7.85cm}

\setlength\figurewidth{.9\defcolwidth}
\setlength\figureheight{.4\figurewidth}

\section{Introduction} \label{sec:intro}

Accurate parameterized feedforward control is essential for the tracking performance of control applications with varying references because of its ability to compensate known disturbances before they affect the system. Approaches to feedforward control range from fully model based methods, such as in \cite{Butterworth2012}, to methods that use data in conjunction with approximate models \citep{VanDeWijdeven2010a}. Through these approaches high performance can be achieved for mechatronic systems including printers \citep{Bolder2014} and wafer stages \citep{Blanken2017c}. 

Iterative learning control (ILC) is a feedforward control approach that can achieve high performance for a single task through complete attenuation of repeating disturbances. In ILC, a feedforward signal is updated iteratively based on the measured error of the previous iteration. Most ILC approaches, including frequency-domain ILC \citep{Bristow2006a} and norm-optimal lifted ILC \citep{Gunnarsson2001}, use measurement data in conjunction with approximate models and as such can improve upon the performance of fully model-based feedforward approaches. However, this performance is only achieved for a single reference, limiting the applicability of these ILC algorithms to systems performing repeating tasks.

The applicability of ILC for industrial applications that perform varying tasks is improved through extensions aimed at increasing the flexibility of ILC. In \cite{Hoelzle2011}, a library of basis tasks is learned that can be combined into references. Basis function approaches to ILC increase the flexibility even further, as shown in, e.g., \cite{VanDeWijdeven2010a,Bolder2014}. In these approaches, feedforward parameters for SISO systems are learned directly and they result in high performance for SISO systems. However, due to the modeling and design requirements the application to MIMO systems with interaction is not trivial, as illustrated in \cite{Blanken2020a}. In addition, since parameterized feedforward control for MIMO systems with interaction involves many interdependent parameters, manual tuning approaches such as described in \cite{Oomen2019} are infeasible in practice. Instead, an automated model-free approach is desired.  

Existing approaches to model-free ILC for MIMO systems with non-repetitive trajectories do not consider basis functions and as such are limited in flexibility. In \cite{Boudjedir2021}, position and velocity-based ILC is applied to a second-order MIMO nonlinear pick-and-place robot, the references for which vary in magnitude and length. Through a model-free gain-based ILC update law performance is improved after a small number of iterations. \cite{Sammons2020} develop a similar approach in which the gain may include system information. While these approaches yield valuable results for certain types of applications, they are limited in the allowed reference variation and consequently are not suitable for mechatronic systems that demand high accuracy for references that show significantly more variation in, e.g., magnitude, velocity and acceleration. For these types of systems and tasks, fully parameterized feedforward is required instead.

Feedforward tuning for MIMO systems through basis function ILC can be framed as an optimization problem aimed at finding feedforward parameters that minimize the tracking error. A similar optimization problem is considered in \cite{Bolder2018}, where non-parameterized MIMO ILC is interpreted as an optimization problem that can be solved without requiring model knowledge. Through experiments on the adjoint system \citep{Ye2005a,Wahlberg2010}, the gradient of a cost function is determined which is then used in a gradient descent ILC algorithm. However, determining the exact gradient requires $n_i \times n_o$ experiments per iteration for a system with $n_i$ inputs and $n_o$ outputs, and as such is not suitable for large MIMO systems. In \cite{Aarnoudse2020b} a different approach is proposed that relates to simultaneous perturbation stochastic approximation (SPSA), see \cite{Spall1988}. This stochastic approximation adjoint ILC approach uses efficient unbiased gradient estimates obtained from a single experiment in a stochastic gradient descent algorithm.  

A different SPSA-related algorithm for finding MIMO feedforward parameters for a FIR basis through gradient approximation is developed in \cite{Heertjes2010a}. Here, both the Hessian and the gradient are estimated from a single simultaneous perturbation experiment and single-trial convergence is assumed. However, as shown in \cite{Spall1997} estimating the Hessian without bias requires additional experiments, and since both the Hessian and the gradient are approximated, multiple iterations are required for convergence in practice. In addition, this approach does not take advantage of the more accurate gradient estimates that are available through measurements on the adjoint system.

Although there exist many results on ILC with basis functions and model-free ILC, a method for efficient model-free tuning of feedforward parameters for MIMO systems is lacking. The aim of this paper is to exploit ideas regarding efficient model-free ILC in the practically relevant setting of automated feedforward tuning for MIMO systems. The contribution consists of the following cornerstones.
\begin{enumerate}
	\item A feedforward parameterization for MIMO systems is proposed that takes into account interaction (Section 2).
	\item A stochastic gradient descent algorithm with optimal step sizes is proposed to find the parameters that minimize the tracking error in terms of the squared 2-norm (Section 3).
	\item It is shown that an unbiased gradient estimate can be obtained from a single experiment regardless of the size of the MIMO system (Section 4).
	\item The approach is illustrated and compared to a deterministic model-free approach in simulation (Section 6).
\end{enumerate}
Additionally, in Section \ref{sec:approach} implementation aspects for motion systems are discussed and conclusions are given in Section \ref{sec:conclusion}.

Earlier results related to this work appeared in \cite{Aarnoudse2020b}, in which stochastic approximation adjoint iterative learning control is developed for massive MIMO systems, and in \cite{Aarnoudse2022} in which a conjugate gradient descent algorithm is proposed to increase the convergence speed. The current paper builds upon these results to develop a framework for efficient model-free tuning of feedforward parameters for MIMO systems, taking advantage of the idea of stochastic approximation adjoint ILC to obtain inexpensive unbiased gradient estimates. Since stochastic approximation adjoint ILC interprets the MIMO ILC problem as an optimization problem, it can be applied almost directly to automated feedforward tuning for feedforward signals that are linear in the parameters. The number of parameters for MIMO feedforward control is typically limited, therefore the feedforward controller can be tuned efficiently and fast through a gradient-based algorithm and modeling is not required. The resulting approach is straightforward and feasible to implement in practice.

\section{Problem formulation} \label{sec:problem}
In this section, the problem of parameterized feedforward control is introduced. First, an example of mass feedforward is used to illustrate the structure of MIMO basis functions. Then, a general structure for the feedforward parameterization for MIMO systems is proposed, constituting the first contribution. 

\begin{figure}[t]
	\centering
	\includegraphics{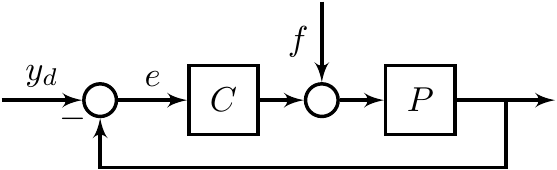}%
	\caption{Closed-loop control system. \label{fig:par_scheme}}%
\end{figure}

Consider a MIMO LTI control system with $n_i$ inputs and $n_o$ outputs as shown in Fig. \ref{fig:par_scheme}, the error $e$ of which is written in lifted form as
\begin{align} \label{eq:error}
	\underbrace{\begin{bmatrix} e^1 \\ \vdots \\ e^{n_o} \end{bmatrix}}_{e} &= \underbrace{\begin{bmatrix} r^1 \\ \vdots \\ r^{n_o} \end{bmatrix}}_{r} - \underbrace{\begin{bmatrix} y^1 \\ \vdots \\ y^{n_o} \end{bmatrix}}_{y}.
	\end{align} 
Here the unknown exogenous disturbance $r=S y_d$, with unknown sensitivity $S=(I+PC)^{-1}$ for plant $P$ and controller $C$, and known reference $y_d$. The output $y$ is given by
\begin{align}
	 \label{eq:output}
	\underbrace{\begin{bmatrix} y^1 \\ \vdots \\ y^{n_o} \end{bmatrix}}_{y} &= \underbrace{\begin{bmatrix} J^{11} & \dots & J^{1 n_i} \\ \vdots & & \vdots \\ J^{n_o 1} & \dots & J^{n_o n_i} \end{bmatrix}}_{J} \underbrace{\begin{bmatrix} f^1 \\ \vdots \\ f^{n_i} \end{bmatrix}}_{f},
\end{align}
with feedforward input $f$ and the unknown process sensitivity $(I+PC)^{-1}P$ denoted by $J$. Here, $e^n,\: r^n,\: y^n,\: f^n \in \mathbb{R}^{N\times 1}$ and $J \in \mathbb{R}^{N n_o \times N n_i}$ for finite signal length $N \in \mathbb{Z}^+$. In this setting both $r$ and $J$ are unknown, $y_d$ and $f$ can be chosen and $e$ can be measured. Note that for $y_d = 0$, $e = Jf$ and therefore we can experiment on $J$. 

\subsection{Example: MIMO mass feedforward}
The aim of parameterized feedforward control is to achieve both high accuracy, i.e., small tracking errors, and flexibility for varying references. To this end, feedforward input $f(y_d,\theta)$ is chosen to be a function of the reference and consists of basis functions $\psi(y_d)$ that are linear in the parameters $\theta$ such that
\begin{align} \label{eq:input}
	f(y_d, \theta) = \psi(y_d) \tran \theta.
\end{align} 
To illustrate a suitable structure for MIMO basis functions for motion control that takes into account interaction, consider the example of mass feedforward for a $2\times 2$ system with reference $y_d = \begin{bmatrix} y_d^1 & y_d^2 \end{bmatrix}\tran$. The acceleration of the reference is taken as basis function, and the resulting feedforward signal is given by
\begin{align}
	\begin{bmatrix} f_m^1(y_d,\theta) \\ f_m^2(y_d,\theta) \end{bmatrix} = \underbrace{\begin{bmatrix}
		\ddot{y_d}^1 & \ddot{y_d}^2 & 0 & 0 \\ 0 & 0 & \ddot{y_d}^1 & \ddot{y_d}^2
	\end{bmatrix}}_{\psi(y_d)\tran} \begin{bmatrix} \theta_1 \\ \theta_2 \\ \theta_3 \\ \theta_4 \end{bmatrix}.
\end{align}
Each of the four transfers in the MIMO system, i.e., both the diagonal and off-diagonal interaction terms, is approximated by a mass line. Feedforward tuning aims to find the mass estimates, the parameters $\theta$, that minimize the tracking error. 

\subsection{General structure for MIMO parameterized feedforward}
A general structure for parameterized feedforward for MIMO systems is given by (\ref{eq:input}), with basis functions structured as 
\begin{align} \label{eq:psi}
	&\psi(y_d)\tran = \\ \nonumber
	& \begin{bmatrix} \psi_1(y_d) & \psi_2(y_d) & \dots & \psi_{n_b}(y_d) & \dots & 0&  0 & \dots & 0 \\ & \vdots  & & & & &  \vdots & & \\ 0 & 0 & \dots & 0 &\dots  &\psi_1(y_d) & \psi_2(y_d) & \dots & \psi_{n_b}(y_d) \end{bmatrix}  
\end{align}
with the $n_b$ basis functions $\psi_n(y_d), \: n = 1,2,...,n_b$ given by
\begin{align}
	 \psi_n(y_d) = \begin{bmatrix}
		\psi^1_n(y_d^1) & \psi^2_n(y_d^2) & \dots & \psi^{n_o}_n(y_d^{n_o})
	\end{bmatrix}.
\end{align} 
Here $\psi^1_n(y_d^1) \in \mathbb{R}^{N\times 1}$ such that $\psi(y_d)^{\tran} \in \mathbb{R}^{n_i N\times n_b n_o n_i}$. The parameter vector is structured as
\begin{align} 	
	\label{eq:theta}
	&\theta = \begin{bmatrix} 
		\theta_1 & \dots & \theta_{n_b n_o n_i}
	\end{bmatrix} \tran \in \mathbb{R}^{n_b n_o n_i\times 1}.
\end{align}
Thus, $\psi(y_d)\tran$ consists of a set of $n_b$ basis function matrices, which is applied to each input direction, $n_i$ times in total. Each of the basis function matrices $\psi_n(y_d)$ consists of a specific basis function applied to each of the $n_o$ output references. The parameter vector $\theta$ contains a separate parameter for each column in each of the $n_i\times n_b$ matrices $\psi_n(y_d)$. This results in $n_i$ feedforward signals of the form
\begin{align} \label{eq:par_input}
	f^n(y_d,\theta) = \sum_{k=1}^{n_o} \sum_{l=1}^{n_b} \psi^k_l(y_d^k) \theta_{(n-1) n_o n_b + (l-1)n_o + k}.
\end{align}
Note that for each input direction, the same set of $n_b$ basis functions is used and only the parameters are different. In addition, to take into account couplings in the MIMO system, the input in each direction contains basis functions based on the reference in all output directions. The framework allows for any type of basis functions that is linear in the parameters, including finite impulse response (FIR) bases or, e.g., non-causal rational basis functions with fixed poles \citep{Blanken2020c}. 

For SISO systems, the parameters $\theta$ can be tuned in different ways, including intuitive tuning by hand or through model-based iterative learning control. For MIMO systems tuning by hand is not trivial since the couplings between system directions may be complex. In addition, it is often difficult to obtain the accurate MIMO models required for model-based ILC. Therefore, the aim of this paper is to develop a method to find the optimal parameters $\theta$ automatically through a series of experiments without requiring model knowledge.

\section{Optimal parameters through stochastic gradient descent} \label{sec:alg}

In this section, the approach to automated feedforward tuning for MIMO systems is introduced. First, the problem of tuning the feedforward parameters is written as an optimization problem in which a cost function is to be minimized. An iterative algorithm is proposed to minimize this cost function through stochastic gradient descent with unbiased gradient estimates and optimal step sizes, leading to the second contribution. Lastly, an overview of the approach is presented.

\subsection{Optimization-based feedforward tuning}
The aim is to find parameters that minimize the tracking error in terms of the squared 2-norm, i.e., to minimize the cost function
\begin{align} \label{eq:cost}
	\mathcal{J}(\theta) = \|e(\theta)\|^2_2,
\end{align}
where the 2-norm is defined as $\|x\|_2 = \sqrt{x\tran x}$, and
\begin{align} \label{eq:e_theta}
	e(\theta) = r - J \psi(y_d)\tran \theta 
\end{align}
according to (\ref{eq:error})-(\ref{eq:input}). The optimal parameters are given by
\begin{align}
	\theta^* = \arg \min_\theta \mathcal{J}(\theta),	
\end{align} 
and since $\mathcal{J}(\theta)$ in (\ref{eq:cost}) is quadratic and convex the parameters can be found through gradient-based optimization. The gradient of (\ref{eq:cost}) is given by
\begin{align}
	g(\theta) &= \frac{\partial \mathcal{J}(\theta)}{\partial \theta} = 2 \psi(y_d) J\tran J \psi(y_d)\tran \theta - 2 \psi(y_d) J\tran r,
\end{align}
which through substitution of (\ref{eq:e_theta}) is rewritten to
\begin{align}
	g(\theta) = -2 \psi(y_d) J\tran e(\theta).
\end{align} 
In Section \ref{sec:gradient} it is shown that an unbiased estimate $\hat{g}(\theta)$ of $g(\theta)$, which includes the transpose of the system $J$ which is unknown in a model-free setting, can be found efficiently using a single experiment regardless of the size of the MIMO system. The gradient estimate is used in a stochastic gradient descent algorithm, in which the parameters are updated iteratively according to
\begin{align} \label{eq:theta_update}
	\theta_{j+1} = \theta_j + \varepsilon_j \hat{g}(\theta_j),
\end{align}
with optimal step size $\varepsilon_j$ in direction $\hat{g}(\theta_j)$ as defined next.

\subsection{Optimal step sizes}

Given a search direction $\hat{g}(\theta_j)$ in the parameter update (\ref{eq:theta_update}), the optimal step size can be found by taking
\begin{align} \nonumber 
	\varepsilon_j &= \arg \min_\varepsilon \mathcal{J} (\theta_{j+1}) \\ \label{eq:cost_step}
	&= \arg \min_{\varepsilon} \|r-J \psi(y_d)\tran (\theta_j + \varepsilon \hat{g}(\theta_j))\|^2_2.
\end{align}

The optimal step size is given in the following theorem.

\begin{thm}
	The optimal step size $\varepsilon_j$ that minimizes (\ref{eq:cost_step}) given a search direction $\hat{g}(\theta_j)$ is given by
	\begin{align} \label{eq:varepsilon}
		\varepsilon_j = \frac{e_j(\theta_j)\tran J \psi(y_d)\tran \hat{g}(\theta_j)}{(J \psi(y_d)\tran \hat{g}(\theta_j))\tran J \psi(y_d)\tran \hat{g}(\theta_j)}.
	\end{align}
\end{thm}

\begin{pf}
	The quadratic criterion (\ref{eq:cost_step}) is minimized by setting
	\begin{align} \label{eq:proof1}
		\frac{\partial \mathcal{J}(\theta_{j+1})}{\partial \varepsilon} = 0.
	\end{align}
	Taking the derivative to $\varepsilon$ gives
	\begin{align} \label{eq:proof2}
		&\frac{\partial}{\partial \varepsilon} \|r-J \psi(y_d)\tran (\theta_j + \varepsilon \hat{g}(\theta_j))\|^2_2 \\ \nonumber
		&= \theta_j\tran \psi(y_d) J\tran J \psi(y_d)\tran \hat{g}(\theta_j) + \hat{g}(\theta_j) \psi(y_d) J\tran J \psi(y_d)\tran \theta_j \\ \nonumber
		&\quad + 2 \varepsilon \hat{g}(\theta_j)\tran \psi(y_d) J\tran J \psi(y_d) \hat{g}(\theta_j) - \hat{g}(\theta_j)\tran \psi(y_d) J\tran r \\ \nonumber &\quad - r\tran J \psi(y_d)\tran \hat{g}(\theta_j).
	\end{align}
	Substituting (\ref{eq:proof2}) in (\ref{eq:proof1}) gives
	\begin{align}
		\varepsilon_j &= \frac{\hat{g}(\theta_j) \psi(y_d) J\tran (r-J \psi(y_d)\tran \theta_j)}{2\hat{g}(\theta_j)\tran \psi(y_d) J\tran J \psi(y_d)\tran \hat{g}(\theta_j)} \\ \nonumber 
		&+ \frac{(r-J \psi(y_d)\tran \theta_j)\tran J \psi(y_d)\tran \hat{g}(\theta_j)}{2\hat{g}(\theta_j)\tran \psi(y_d) J\tran J \psi(y_d)\tran \hat{g}(\theta_j)}. 
	\end{align}
	Substitution of (\ref{eq:e_theta}) and rewriting leads to
	\begin{align}
		\varepsilon_j = \frac{e_j(\theta_j)\tran J \psi(y_d)\tran \hat{g}(\theta_j)}{(J \psi(y_d)\tran \hat{g}(\theta_j))\tran J \psi(y_d)\tran \hat{g}(\theta_j)},
	\end{align}
which concludes the proof.
\end{pf} 

The optimal step size is determined without requiring any model knowledge. Given $e_j(\theta_j)\tran$ and $\hat{g}(\theta_j)$, $\varepsilon_j$ in (\ref{eq:varepsilon}) follows from a single experiment on the system $J$. The experiment with reference $y_d = 0$ takes as feedforward input $\psi(y_d)\tran \hat{g}(\theta_j)$, resulting in the output $J \psi(y_d)\tran \hat{g}(\theta_j)$ from which both the numerator and the denominator term of (\ref{eq:varepsilon}) are determined.

\subsection{Overview of the approach}

The unbiased gradient estimate in Section \ref{sec:gradient} is combined with the optimal step size in (\ref{eq:varepsilon}) to obtain parameter update $\theta_{j+1}$ according to (\ref{eq:theta_update}). In addition to the experiment to determine the error $e_j(\theta_j)$ of iteration $j$, the parameter update requires only two experiments to determine the optimal step size and gradient estimate, regardless of the size of the MIMO system. The complete approach is summarized in Algorithm \ref{alg:approach}.

\begin{algorithm}[H]
	\caption{Model-free automated feedforward tuning} 	\label{alg:approach}
	\begin{algorithmic}[1]
		\State{\textbf{for} $j=1:n_{\text{iteration}}$}
		\State{\quad Apply input $f_j = \psi(y_d)\tran \theta_j$ and measure }
		\Statex{\quad $e_j(\theta_j) = r - J \psi(y_d)\tran \theta_j$.}
		\State{\quad Find unbiased estimate $\hat{g}(\theta_j)$ using one experiment}
		\Statex{\quad   according to Section \ref{sec:gradient}, Eq. (\ref{eq:grad_bar}).}
		\State{\quad Measure $J \psi(y_d)\tran \hat{g}(\theta_j)$ to find step size $\varepsilon_j$ in (\ref{eq:varepsilon}).}	
		\State{\quad Update $\theta_{j+1} = \theta_j - \varepsilon_j \hat{g}(\theta_j)$.}		
		\State{\textbf{end}}	
	\end{algorithmic}
\end{algorithm}

\section{Unbiased gradient estimates} \label{sec:gradient} 

The approach introduced in Section \ref{sec:alg} uses unbiased estimates of the gradient of (\ref{eq:cost}) for automated tuning of the feedforward parameters. In this section, a single experiment on system $J$ is used to obtain these estimates, leading to the third contribution. 

The gradient at the point $\theta_j$ is given by
\begin{align} \label{eq:grad}
	g(\theta_j) = \frac{\partial \mathcal{J}}{\partial \theta}(\theta_j) = -2 \psi(y_d) J\tran e_j(\theta_j),
\end{align}
with $J\tran$ the adjoint operator of $J$, which is defined as follows.

\begin{defn}
	Let $\langle f,g \rangle = f\tran g$ denote the inner product of two signals $f,g \in \mathbb{R}^{N \times 1}$. The adjoint $J^*$ of $J$ is defined as the operator that satisfies the condition
	\begin{align} \nonumber
		\langle f,Jg \rangle = \langle J^* f, g \rangle, \quad \forall f,g \in \mathbb{R}^{N\times 1}.
	\end{align}
	The adjoint $J^*$ of $J$ is given by $J\tran$, which follows from
	\begin{align} \nonumber
		f^\top Jg = (J^* f)\tran g = f\tran (J^*)\tran g, \quad \forall f,g \in \mathbb{R}^{N \times 1}.
	\end{align}
\end{defn}

The adjoint operator relates to $J$ through a time reversal, resulting in the following lemma for SISO LTI systems. 
\begin{lem} \label{lem:adj_siso}
	The adjoint of a SISO LTI system $J = J^{11}$ is given by $\left(J^{11}\right)\tran = \mathcal{T} J^{11} \mathcal{T}$, where the involutory permutation matrix 
	\begin{align*} 
		\mathcal{T} = \begin{bmatrix}
			0 & \dots &  0 & 1 \\
			\vdots &  &  1 & 0 \\
			0 & \iddots &  & \vdots \\
			1 & 0 & \dots & 0  
		\end{bmatrix} \in \mathbb{R}^{N \times N}
	\end{align*}
	has the interpretation of a time-reversal operator. 
\end{lem}

For SISO systems, Lemma \ref{lem:adj_siso} enables direct experiments on $J\tran$ through a single experiment on $J$ with two time reversals. For non-symmetric MIMO systems this is not applicable, as follows from the following lemma.

\begin{lem} \label{lem:adj_mimo}
	The adjoint of a MIMO LTI system $J$ is given by
	\begin{align} \label{eq:adj_mimo} 
		J\tran &=  \begin{bmatrix} (J^{11})\tran & \dots & (J^{n_o 1})\tran \\ \vdots & & \vdots \\ (J^{1 n_i})\tran & \dots & (J^{n_o n_i})\tran \end{bmatrix}   \\ \nonumber
		&=\underbrace{\begin{bmatrix} \mathcal{T} & & 0 \\ & \ddots & \\ 0 & & \mathcal{T} \end{bmatrix}}_{\mathcal{T}^{n_i}} 
		\underbrace{\begin{bmatrix} J^{11} & \dots & J^{n_o 1} \\ \vdots & & \vdots \\ J^{1 n_i} & \dots & J^{n_o n_i} \end{bmatrix}}_{\tilde{J}} \underbrace{\begin{bmatrix} \mathcal{T} & & 0 \\ & \ddots & \\ 0 & & \mathcal{T} \end{bmatrix}}_{\mathcal{T}^{n_o}}.
	\end{align}
\end{lem}
For non-symmetric MIMO systems, $\tilde{J} \neq J$, such that the term $J \tran e_j$ in (\ref{eq:grad}) cannot be determined from a single experiment on $J$. However, an unbiased estimate $\hat{g}(\theta_j)$ can be determined from a single experiment according to the following theorem.

\begin{thm}
	An unbiased estimate $\hat{g}(\theta_j)$ of (\ref{eq:grad}) is given by
	\begin{align} \label{eq:grad_bar}
		\hat{g}(\theta_j) = -2 \psi(y_d) \mathcal{T}^{n_i} A_j J A_j \mathcal{T}^{n_o} e_j(\theta_j).
	\end{align}
	The matrix $A_j \in \mathbb{R}^{(N n_i) \times (N n_o)}$ is given by
	\begin{align} \label{eq:A} 
		A_j = \begin{bmatrix}
			a^{11}_j & \dots & a^{1n_o}_j\\
			\vdots & \ddots &  \vdots \\
			a^{n_i 1}_j & \dots & a^{n_i n_o}_j 
		\end{bmatrix} \otimes I_N
	\end{align}
	where $I_N$ is the $N\times N$ identity matrix and the entries $a^{lm}_j$ are samples from a symmetric Bernoulli $\pm 1$ distribution, i.e., $a^{lm}_j\in \{-1,1\}$ and the probabilities are given by $P(a^{lm}_j = 1) = 1/2$ and $P(a^{lm}_j = -1) = 1/2$. 
\end{thm}

\begin{pf}
	It holds that
	\begin{align} \label{eq:25}
		\mathcal{T}^{n_i} A_j J A_j \mathcal{T}^{n_o} e_j(\theta_j) = J\tran e_j(\theta_j) + \eta_j,
	\end{align} 
	where $\eta_j$ is interpreted as an unbiased disturbance term, i.e.,
	\begin{align}
		\mathbb{E}\{\eta_j\} = 0,
	\end{align}
	and therefore 
	\begin{align}
		\mathbb{E}\{\mathcal{T}^{n_i} A_j J A_j \mathcal{T}^{n_o} e_j(\theta_j)\} = J\tran e_j(\theta_j),
	\end{align}
	according to the proof of \cite[Theorem 1]{Aarnoudse2020b}.	Substituting unbiased estimate (\ref{eq:25}) in (\ref{eq:grad}) gives
	\begin{align}
		\hat{g}(\theta_j)&= -2 \psi(y_d)  J\tran e_j(\theta_j) - 2 \psi(y_d) \eta_j.
	\end{align} 
	Since $\mathbb{E}\{\eta_j\} = 0$, it follows from the linearity of the expected value operator that $\mathbb{E}\{\psi(y_d) \eta_j\} = 0$. Therefore,
	\begin{align}
		\mathbb{E}\{\hat{g}(\theta_j)\} = -2 \psi(y_d)  J\tran e_j(\theta_j) = g(\theta_j)
	\end{align}
	which concludes the proof.	
\end{pf} 

Thus, a single experiment gives an unbiased gradient estimate.

\section{Implementation for motion systems} \label{sec:approach}

In this section, suitable basis functions for MIMO motion systems are given and some implementation aspects are discussed.

\subsection{Basis functions}

Parameterized feedforward control aims to minimize the error for any reference. From (\ref{eq:error}) and (\ref{eq:output}) it follows that zero tracking error requires 
\begin{align}
	f = J^{-1} r = (SP)^{-1} S y_d = P^{-1} y_d.
\end{align}
Therefore, the basis functions that construct the feedforward signal should approximate the inverse plant well. For typical motion systems, suitable basis functions are the position reference and its derivatives: velocity, acceleration and snap terms compensate respectively viscous friction, mass dynamics and the compliance of the flexible dynamics \citep{Oomen2019, Bolder2014}. This leads to the following basis functions.
\begin{align} \label{eq:psi2}
	&\psi(y_d)\tran = \begin{bmatrix} y_{d,\tau} & y_{d,\tau}' & \dots & y_{d,\tau}^{(4)} & \dots & 0&  0 & \dots & 0 \\ & \vdots  & & & & &  \vdots & & \\ 0 & 0 & \dots & 0 &\dots  &y_{d,\tau} & y_{d,\tau}' & \dots & y_{d,\tau}^{(4)} \end{bmatrix}  
\end{align}
in which
\begin{align}
	y_{d,\tau} = \begin{bmatrix}
		y_d^1 & y_d^2 & \dots y_d^{n_o}
	\end{bmatrix}.
\end{align}
This parameterization, which relates each of the output references to each input direction, models both the diagonal and off-diagonal terms of the MIMO system and is therefore suitable for MIMO systems with strong interaction.

\subsection{Implementation considerations}
Some practical considerations are of importance for the implementation of the experiments to obtain the gradient estimate and the step size. Many motion systems show stick-slip friction due to which small input signals may not result in useful outputs. When small error and gradient signals are used as inputs, their magnitude may be too small to push the system outside of the stick regime. This is solved by using scaling factors $\alpha_j \in \mathbb{R}$ and $\beta_j\in \mathbb{R}$, implemented as
\begin{align}
	\hat{g}(\theta_j) &= -2 \psi(y_d) \mathcal{T}^{n_i} A_j \frac{1}{\alpha_j} J \alpha_j  A_j \mathcal{T}^{n_o} e_j(\theta_j), \\ 
	J \psi(y_d)\tran \hat{g}(\theta_j) &= \frac{1}{\beta_j}J \beta_j \psi(y_d)\tran \hat{g}(\theta_j).
\end{align}
It is assumed that $\alpha_j$ and $\beta_j$ are chosen such that the system is outside of the stick regime during the experiment, and that it therefore behaves linearly. Since the units of feedforward and error signals typically differ, $\alpha_j$ may also be used to reduce the magnitude of the input for the gradient estimation experiment.

\section{Simulation example} \label{sec:example}

The proposed method is illustrated using an industrial flatbed printer in simulation. First, the setup is introduced. Then, simulation results are given and lastly, the stochastic approach is compared to a deterministic approach that uses exact gradients.

\subsection{Setup}
\begin{figure}[t]
	\centering
	\includegraphics[width=0.8\linewidth]{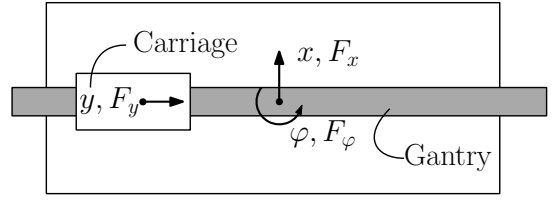}
	\caption{Schematic representation of the Arizona flatbed printer. \label{fig:arizona}}
\end{figure}

\begin{figure}[t]	
	\centering
	\setlength\figurewidth{.9\defcolwidth}
	\setlength\figureheight{.9\figurewidth}
	\includegraphics{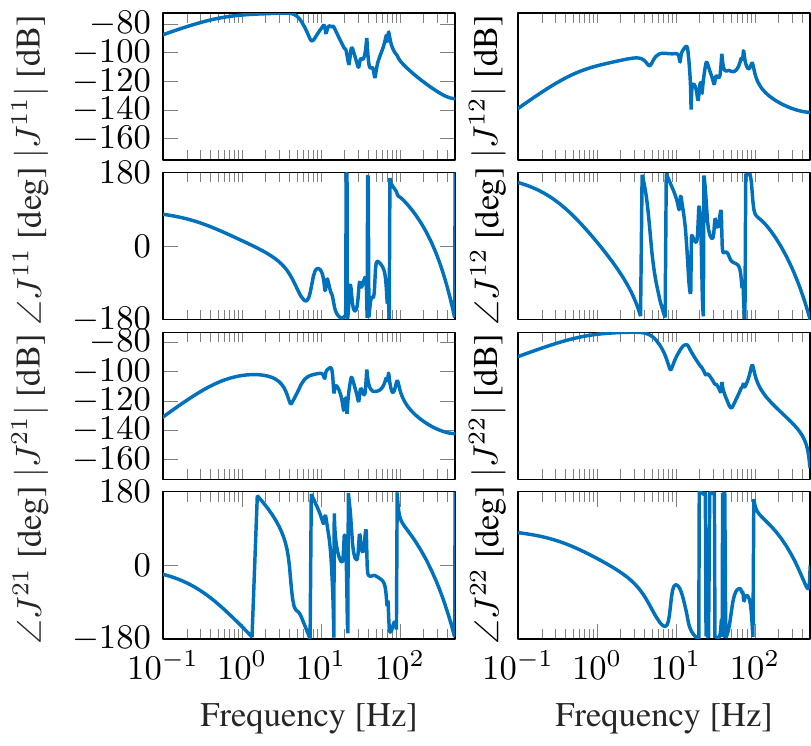}
	\caption{Bode diagram of the transfer function $J = (I+PC)^{-1}P$ for the gantry of the Arizona flatbed printer.}
	\label{fig:bode}
\end{figure}

\begin{figure}[t]
	\centering
	\setlength\figurewidth{.9\defcolwidth}
	\setlength\figureheight{.4\figurewidth}
	\includegraphics{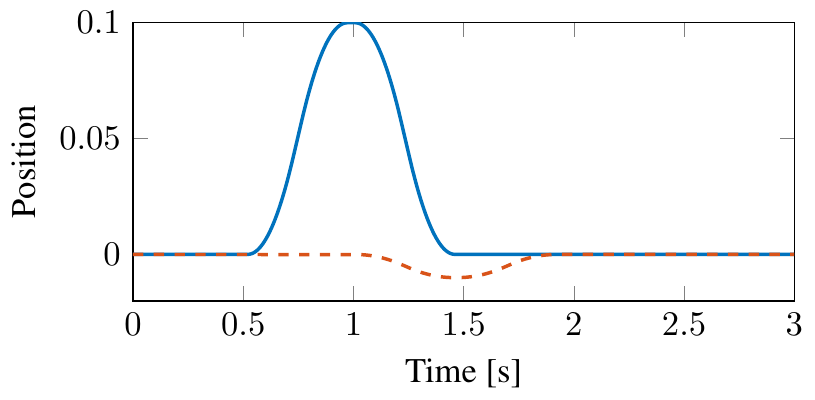}
	\caption{Reference consisting of a translation in $x$-direction (\protect\blueline, [\si{\meter}]) and a small rotation in $\varphi$-direction (\protect\reddash, [\si{\radian}]) \label{fig:refs}}
\end{figure}

The approach is illustrated using a simulated Arizona flatbed printer, schematically represented in Fig. \ref{fig:arizona}. In the simulation a $2\times 2$ version of the system is used with force input $\begin{bmatrix} f^1 & f^2 \end{bmatrix}\tran = \begin{bmatrix} F_x & F_\varphi \end{bmatrix}\tran$ and position output $\begin{bmatrix} y^1 & y^2 \end{bmatrix}\tran = \begin{bmatrix} x & \varphi \end{bmatrix}\tran$. A Bode diagram of the process sensitivity $J = (I+PC)^{-1}P$ is shown in Fig. \ref{fig:bode}. The fourth-order reference $y_d$ consists of a translation and a small rotation of the gantry as shown in Fig. \ref{fig:refs}.

\subsection{Results}

Model-free automated feedforward tuning is applied to the Arizona printer in simulation. The basis functions are chosen, according to Section \ref{sec:approach}, as the position reference, velocity, acceleration, jerk and snap such that $n_b = 5$. In accordance with Algorithm \ref{alg:approach}, each iteration requires three experiments to determine respectively the error, the unbiased gradient estimate and the step size. For the model-free algorithm the error converges within five iterations, requiring fifteen experiments in total, see Fig. \ref{fig:conv_comp}. 

\subsection{Comparison to a deterministic model-free approach}

\begin{figure}[t]
	\centering
	\setlength\figurewidth{.9\defcolwidth}
	\setlength\figureheight{.4\figurewidth}
	\includegraphics{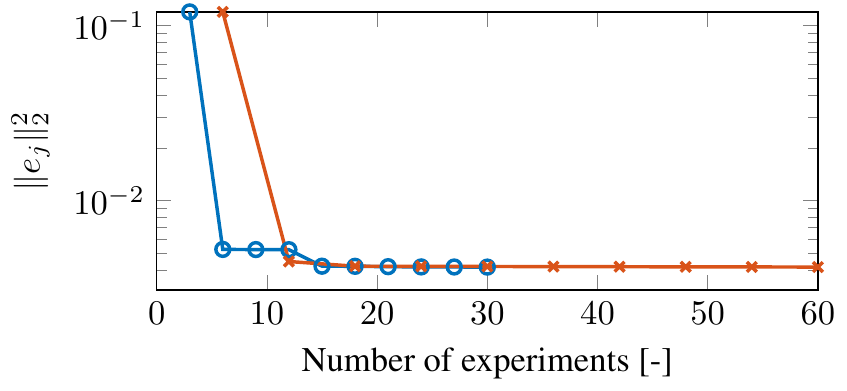}
	\caption{Stochastic approximation adjoint ILC for MIMO feedforward tuning (\protect\bluedot) converges within 15 experiments (for five iterations) while the deterministic approach ($\textcolor{matred1}{\times}$) requires 18 experiments (for three iterations) to converge. \label{fig:conv_comp}}
\end{figure}

An alternative model-free approach uses exact gradient expressions that require $n_i \times n_o$ experiments to obtain according to
\begin{align} 
	g(\theta_j) = -2 \psi(y_d) \mathcal{T}^{n_i} \left( \sum_{n_i}^{l=1} \sum_{n_o}^{m=1} E^{lm} J E^{lm} \right) \mathcal{T}^{n_o} e_j(\theta_j),
\end{align}
where $E^{lm}$ consists of zeros, with a one on the $lm^{th}$ entry, see also \cite{Bolder2018} where this is applied to norm-optimal ILC. This deterministic approach scales badly for large MIMO systems due to the $n_i \times n_o$ experiments required per iteration to determine the gradient \citep{Aarnoudse2020b}. In Fig. \ref{fig:conv_comp} it is shown that even for a $2 \times 2$ system, the stochastic approach requires fewer experiments to converge.

\section{Conclusion} \label{sec:conclusion}

In this paper a new approach is developed for model-free automated feedforward tuning for MIMO systems that uses efficient unbiased gradient estimates in a stochastic gradient descent optimization algorithm. Through specific pre- and post-multiplications, the estimate of the gradient of a cost criterion is determined from a single experiment regardless of the size of the MIMO system. An additional experiment is used to determine the optimal step size in the direction of the gradient estimate, resulting in fast convergence. The approach is illustrated in a simulation example, where it is shown to require fewer experiments to converge than a deterministic approach that uses exact gradients, even for a relatively small $2\times 2$ system. Since the number of parameters in feedforward tuning is limited, the required number of iterations is small such that the approach is feasible in practice. Future developments involve experimental implementation of the proposed approach and extension of the set of basis functions to further increase performance.

\begin{ack}
The authors wish to acknowledge Shaun Boyteen Joseph for the discussions on the topic which have contributed to this result.
\end{ack}

\bibliography{library}


\end{document}